\documentclass[prb,preprint,showpacs,preprintnumbers,amsmath,amssymb]{revtex4}

\usepackage[dvipdfmx]{graphicx}
\usepackage{bm}

\newcommand{\be}{\begin{equation}}
\newcommand{\ee}{\end{equation}}
\newcommand{\eq}[1]{Eq.~(\ref{#1})}
\newcommand{\fig}[1]{Fig.~\ref{#1}}
\def\bea{\begin{eqnarray}}
\def\eea{\end{eqnarray}}

\def\bra{\langle}
\def\ket{\rangle}
\def\vq{{\bf q}}

\def\vk{{\bf k}}
\def\vQ{{\bf Q}}

\begin{document}

\title{Suppression of superconductivity by spin fluctuations in iron-based superconductors}

\author{Hiroyuki Yamase$^{1,2,3}$ and Tomoaki Agatsuma$^{1,2}$}
\affiliation{
{$^{1}$}National Institute for Materials Science, Tsukuba 305-0047, Japan \\
{$^{2}$}Department of Condensed Matter Physics, Graduate School of Science, 
Hokkaido University, Sapporo 060-0810, Japan \\
{$^{3}$}International Center for Materials Nanoarchitectonics, National Institute for Materials Science, Tsukuba 305-0003, Japan
}

\date{November 20, 2019}

\begin{abstract}
We study the superconducting instability mediated by spin fluctuations 
in the Eliashberg theory for a minimal two-band model of iron-based superconductors. 
While antiferromagnetic spin fluctuations can drive superconductivity (SC) as is well established, 
we find that spin fluctuations necessarily contain a contribution to suppress SC 
even though SC can eventually occur at lower temperatures. 
This self-restraint effect stems from a general feature of the spin-fluctuation mechanism, 
namely the repulsive pairing interaction, which leads to 
phase frustration of the pairing gap and consequently the suppression of SC. 
\end{abstract}


\maketitle
Iron-based superconductors (FeSC) provide a platform to explore a mechanism of 
high-temperature (high-$T_c$) superconductivity (SC) \cite{stewart11}. 
Since SC is realized close to a spin-density-wave (SDW) phase, 
the importance of spin fluctuations is widely recognized as 
a possible mechanism of SC \cite{mazin08,kuroki08,chubukov08}.  
A close look at the phase diagram of FeSC reveals the presence of an electronic nematic phase, 
which is also close to the SC phase. 
While the origin of the nematic phase is still controversial \cite{fernandes14}, 
it was shown that orbital nematic fluctuations lead to strong coupling SC 
with an onset temperature comparable to the observation \cite{yamase13b,agatsuma16}. 
The electronic structure of FeSC 
is characterized by multibands originating from five $3d$ orbitals of Fe ions \cite{kuroki08}. 
Hence the orbital fluctuations are also explored 
as a possible mechanism of SC \cite{stanescu08,kontani11,misc-orbital}. 
While electron-phonon coupling is present in real materials and is expected to 
lead to SC, the transition temperature ($T_c$) is believed to be too low compared to 
the observation \cite{boeri08}. 

The distinction between different SC mechanisms is a key issue of FeSC. 
Typically spin fluctuations lead to the so-called $s_{\pm}$-wave symmetry \cite{mazin08,kuroki08,chubukov08} 
whereas  nematic \cite{yanagi10,yamase13b,agatsuma16} 
and orbital \cite{stanescu08,kontani11} fluctuations yield 
$s_{++}$-wave symmetry. 
Obviously this symmetry difference is crucial, but it is not easy to resolve the phase of SC order 
in experiments. 
Furthermore, an $s_{\pm}$-wave pairing gap was found to be stabilized even for nematic fluctuations 
when a partial contribution from spin fluctuations is considered in Ref.~\onlinecite{yamada14}, suggesting 
that the gap symmetry itself cannot be decisive in identifying  the SC mechanism. 

The momentum dependence of the pairing gap is expected to depend on the underlying 
SC mechanism. However, it turned out \cite{agatsuma16} that nematic fluctuations 
lead to a pairing gap similar to that from spin fluctuations, except for the sign of the pairing gap. 
Considering simplifications involved in many theoretical studies, it is not easy to extract 
a robust and key difference of the gap structure, which can distinguish between 
the different SC mechanisms.

Irrespective of the underlying SC mechanism in FeSC, it is tacitly assumed 
that spin, orbital, and nematic fluctuations work positively on driving SC. 
However, in this paper, we find that spin fluctuations tend to suppress 
the SC instability even though spin fluctuations can eventually lead to SC at lower temperatures. 
This self-restraint effect 
is a general feature originating from a repulsive pairing interaction, which yields a sign change of the pairing gap 
on the Fermi surfaces (FSs) connected by a momentum transfer of the spin fluctuations. 

A minimal model for the band structure of FeSC may read as \cite{raghu08,yao09} 
\be
H_{0}= \sum_{\vk,\sigma,\alpha,\beta} \epsilon_{\vk}^{\alpha \beta} 
c_{\vk \alpha \sigma}^{\dagger} c_{\vk \beta \sigma} \,
\label{H0}
\ee
on a square lattice, where the unit  cell contains one iron and $\alpha=1$ and $2$ refer to 
the $d_{xz}$ and $d_{yz}$ orbital, respectively; $c_{\vk \alpha \sigma}^{\dagger}$ and 
$c_{\vk \alpha \sigma}$ are the creation and annihilation operators for electrons 
with momentum $\vk$, orbital $\alpha$, and spin orientation $\sigma$; 
intraorbital dispersions are given by 
$\epsilon^{11}_{\bf k} = -2t_1 \cos k_x -2t_2 \cos k_y -4t_3 \cos k_x 
\cos k_y -\mu$ and 
 $\epsilon^{22}_{\bf k} = -2t_2 \cos k_x -2t_1 \cos k_y -4t_3 \cos k_x 
\cos k_y - \mu$, whereas the the interorbital dispersion is $\epsilon^{12}_{\bf k} = -4t_4 \sin k_x \sin k_y$; 
$\mu$ is the chemical potential. 
The typical FSs observed  in FeSC are well captured by choosing the parameters  
as \cite{yao09} 
$t=-t_1$, $t_2/t=1.5$, $t_3/t=-1.2$, $t_4/t=-0.95$, and $\mu/t=0.6$. 
In the following, we measure all quantities with the dimension of energy in units of $t$. 

\begin{figure}[ht]
\centering
\includegraphics[width=7cm]{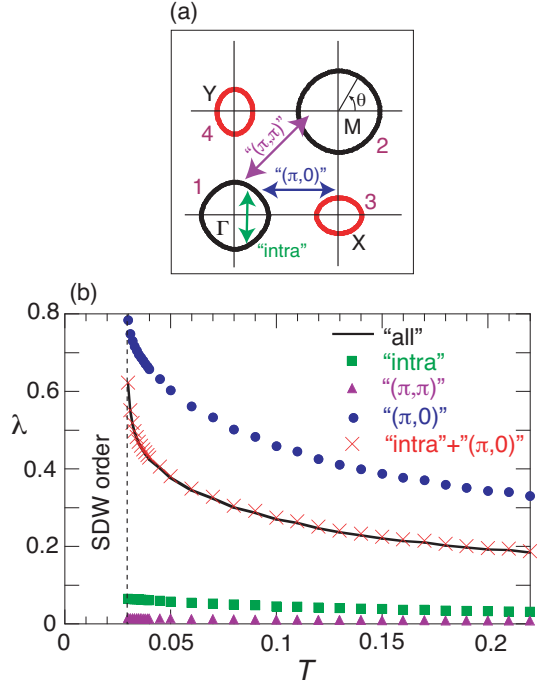}
\caption{(Color online) 
(a) Hole Fermi pockets (1 and 2) around $\Gamma$ and $M$ points 
and electron pockets (3 and 4) around $X$ and $Y$ in the normal state. 
"intra", $"(\pi,0)"$, and $"(\pi,\pi)"$ denote 
scattering processes inside each pocket, between the hole and electron pockets, 
and between the two hole (or electron) pockets, respectively. 
(b) Temperature dependence of the eigenvalues $\lambda$ (solid line). 
The eigenvalues are also computed by focusing on 
particular scattering processes as denoted by 
"intra", "$(\pi,\pi)$", and "$(\pi,0)$". 
Below $T=0.030$, SDW order occurs before SC instability. 
}
\label{lambda-T}
\end{figure}

As shown in \fig{lambda-T}~(a), the Hamiltonian (\ref{H0}) 
yields two hole FSs around $\vk=(0,0)$ and $(\pi,\pi)$, and 
two electron FSs around $\vk=(\pi,0)$ and $(0,\pi)$, which we refer to as 
FS1, FS2, FS3, and FS4, respectively. 
FS1 and FS2 originate from both $d_{xz}$ and $d_{yz}$ orbitals whereas 
FS3 consists of $d_{yz}$ orbital and FS4 $d_{xz}$ orbital. 
These FSs capture the orbital components obtained in a more realistic 5-band model \cite{graser09}. 

To clarify the effect of spin fluctuations on SC, 
we consider a general SU(2) symmetric two-particle interaction 
\bea
&&H_{I}= \frac{1}{8N} \sum_{\vq,\vk,\vk'}\sum_{\alpha, \beta,\sigma_j}
V(\vk,\vk',\vq) \times  \nonumber \\
&&\hspace{5mm}  
{\boldsymbol \sigma}_{\sigma_1 \sigma_2} \cdot  {\boldsymbol \sigma}_{\sigma_3 \sigma_4} 
c_{\vk \alpha \sigma_1}^{\dagger} c_{\vk+\vq \alpha \sigma_2} 
c_{\vk'+\vq \beta \sigma_3}^{\dagger} c_{\vk' \beta \sigma_4} 
\label{HI}
\eea
where $j$ runs from 1 to 4, $\boldsymbol{\sigma}$ are Pauli matrices, and 
$N$ is the total number of the lattice sites. 
This interaction is the effective one close to a SDW phase. 
It should not be associated with the Heisenberg-type spin interaction in the strong coupling physics,  
because our model is defined in the usual Hilbert space where the double occupancy of electrons 
is allowed at any site. Microscopically the interaction (\ref{HI}) is obtained 
as a low-energy effective magnetic interaction generated by, 
for example, the repulsive Hubbard interaction by decreasing the energy scale 
in a functional renormalization group scheme \cite{husemann09,eberlein14}. 
The form of $V(\vk,\vk',\vq)$ depends on details of high-energy fluctuations. 
To keep a connection with FeSC, we approximate $V(\vk, \vk', \vq) \approx V(\vq)$, 
so that a conventional SDW order can be stabilized. $V(\vq)$  should 
exhibit a peak at $\vq=(\pm \pi,0)$ and $(0, \pm\pi)$ with a negative sign 
to capture the stripe-type antiferromagnetic order typically observed in FeSC \cite{dai12}. 
We consider $V(\vq) = 2V_1 (\cos q_x + \cos q_y) + 4 V_2 \cos q_x \cos q_y$ 
with $V_2 > V_1 /2 >0$; we put $V_1=1$ for simplicity.  
In this case, the sizable interaction extends up to the second nearest-neighbor sites in real space. 
One may consider a different form of $V(\vq)$, but  
our major conclusions do not change \cite{misc-supp-Lorentz}. 

For the interaction described by \eq{HI}, the spin fluctuation propagator is computed from 
a bubble summation, namely 
\be
\tilde{V}(\vq, i q_m) = V(\vq) - \frac{V(\vq) \chi_{0}(\vq, i q_m) V(\vq)}{1 +  V(\vq) \chi_{0}(\vq, i q_m)}
\label{Vtilde}
\ee
and $\chi_{0} (\vq, i q_m) = - \frac{T}{2N} \sum_{\vk, \sigma, n} {\rm Tr} \mathcal{G}_{0} (\vk, i k_n) \mathcal{G}_{0}(\vk+\vq, i k_n+ i q_m)$. Here $\mathcal{G}_{0}$ is a $2 \times 2$ matrix of the noninteracting 
Green function defined for \eq{H0}, $i k_n$ ($i q_m$) fermionic (bosonic) Matsubara frequency, and $T$ temperature. 
The first term in \eq{Vtilde} does not depend on frequency and describes the instantaneous effect, 
whereas the second term accounts for the retardation effect on the pairing. 
A role of the instantaneous part for SC would be analyzed appropriately by including the Coulomb 
repulsion \cite{schrieffer}. As a result, the superconducting tendency from the instantaneous part 
would be significantly suppressed. Even in this case, 
as we shall show below, the self-restraint effect itself is general and can occur also 
for the instantaneous part as long as it provides the repulsive pairing interaction. 
However, we believe that the dynamical effect is more important than the instantaneous effect 
as widely discussed for FeSC. 
To make the new mechanism of the self-restraint effect transparent 
as much as possible, we focus on dynamical spin fluctuations described by the second term in \eq{Vtilde}. 

The Eliashberg gap equations involve two coupled nonlinear equations for the pairing gap 
$\Delta (\vk, i k_n)$ and the renormalization function $Z(\vk,i k_n)$. 
In many interesting cases, it is highly demanding to solve the Eliashberg equations 
numerically. Hence $Z(\vk, i k_n)$ would be set to unity and yet computation would be 
limited to a temperature region much higher than $T_c$. 
To overcome these technological issues, we recall that SC instability is a phenomenon 
close to the FS and project the momentum on the FSs. 
We divide the FSs into many patches and define the Fermi momentum $\vk_F$ on each patch. 
Thus $\vk_F$ is a discrete quantity in this work. This idea allows us to achieve 
stable computations down to very low temperature 
with including the renormalization function \cite{yamase13b} as well as a fine 
momentum resolution \cite{agatsuma16}. 

After linearizing the Eliashberg equations with respect to $\Delta(\vk, ik_n)$, 
we obtain 
\bea
&&\Delta(\vk_F, i k_n) Z(\vk_F, i k_n) = \nonumber \\ 
&&\hspace{10mm}  - \pi T \sum_{\vk_{F}', n'} N_{\vk_{F}'} \frac{\Gamma_{\vk_{F} \vk_{F}'} (i k_n, ik_{n}')}{| k_{n}' |} \Delta (\vk_{F}', i k_{n}') \,,
\label{gap-eq} \\
&& Z(\vk_F, i k_n) = 1- \pi T \sum_{\vk_{F}', n'} N_{\vk_{F}'} \frac{k_{n}'}{k_n} \frac{\Gamma_{\vk_{F}\vk_{F}'}^{Z}  (i k_n, ik_{n}')} {| k_{n}' |} \, .
\label{Z-eq}
\eea
Here $N_{\vk_{F}}$ is a momentum-resolved density of states on each FS patch and 
$\Gamma_{\vk_{F} \vk_{F}'} (i k_n, i k_{n}')$ is the averaged pairing interaction over the FS patches 
specified by $\vk_{F}$ and $\vk_{F}'$: 
\bea
&&\Gamma_{\vk_F \vk_{F}'}(i k_n, i k_n') = - \frac{1}{4} \left\bra W_{a b}(\vk, \vk')^2 \times
\right. \nonumber \\
&&\hspace{1mm} 
\left. 
\left( \tilde{V}(\vk-\vk', i k_n - i k_n') + 2 \tilde{V} (\vk+\vk', i k_n + i k_n') \right) \right\ket_{\vk_F \vk_{F}'} \, ,
\label{averagedGamma}
\eea
where $\tilde{V}(\vk-\vk', i k_n - i k_n')$ comes from longitudinal spin fluctuations 
and $\tilde{V} (\vk+\vk', i k_n + i k_n')$ transverse ones. 
The vertex part $W_{a b}(\vk, \vk') = \left( U^{\dagger}(\vk) U(\vk') \right)_{a b}$ comes 
from the $2 \times 2$ unitary matrix diagonalizing the kinetic term \eq{H0}, and $a$ and $b$ 
denote band indices. Since each band forms FSs, the indices $a$ and $b$ can be absorbed 
into the FS indices $\vk_{F}$ and $\vk_{F}'$. 
Similarly, we can compute $\Gamma_{\vk_F \vk_{F}'}^{Z}(i k_n, i k_n')$ in \eq{Z-eq} as 
\bea
&&\Gamma_{\vk_F \vk_{F}'}^{Z}(i k_n, i k_n') = \frac{1}{4} \left\bra W_{a b}(\vk, \vk')^2 \times 
\right.  \nonumber \\
&& \hspace{10mm} 
\left. \left( 3 \tilde{V}(\vk-\vk', i k_n - i k_n') - 2 V (\vk - \vk') \right) \right\ket_{\vk_F \vk_{F}'} \, .
\label{averagedGammaZ}
\eea
$Z(\vk_F, i k_n)$ is directly obtained from \eq{Z-eq}. It is then straightforward to solve the 
eigenvalue equation 
\eq{gap-eq} numerically. When the eigenvalue $\lambda$ exceeds unity, SC instability occurs.

Since the SC instability is expected near the antiferromagnetic phase, we choose $V_2 = 1.7$, 
for which the stripe-type SDW order occurs below $T=0.030$. 
The value of $V_2$ is a control parameter to tune the SDW phase in our low-energy effective model 
and our conclusion of the self-restraint effect does not depend on a choice of $V_2$. 

\begin{figure}[th]
\centering
\includegraphics[width=7cm]{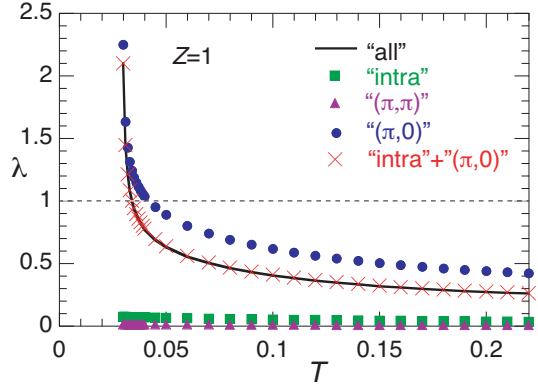}
\caption{(Color online) 
Temperature dependence of the eigenvalues $\lambda$ (solid line denoted by "all"). 
The eigenvalues are also computed by focusing on 
particular scattering processes as denoted by 
"intra", "$(\pi,\pi)$", and "$(\pi,0)$". 
The self-energy effect is discarded by assuming $Z=1$. 
}
\label{lambda-T-Z1}
\end{figure}

The solid line in \fig{lambda-T} (b) shows  the temperature dependence of 
the eigenvalue of \eq{gap-eq}. 
With decreasing temperature, the eigenvalue is enhanced and reaches as large as 
0.6 at $T \approx 0.03$. 
If the temperature is decreased further, SDW instability would preempt SC instability. 
While the SC instability therefore does not occur in a strict sense, 
the eigenvalue less than unity 
is frequently obtained in many theoretical studies for FeSC  and 
consistent with the literature \cite{arita09,suzuki14,usui15}. 
Note that the eigenvalue can exceed unity if we neglect the self-energy effect (see \fig{lambda-T-Z1}).

For the FSs typical to FeSC, there are three different low-energy scattering processes 
"intra", "$(\pi,0)$", and "$(\pi,\pi)$" as shown in \fig{lambda-T}~(a). 
To identify the dominant  scattering process leading to the SC, 
we also compute the eigenvalue of the Eliashberg 
equation \eq{gap-eq} by choosing particular scattering processes. 
Since spin fluctuations are characterized by momenta $(\pi,0)$ and $(0,\pi)$, 
it is reasonable that the eigenvalue for "$(\pi,0)$" scattering processes becomes much larger than the other two. 
Our finding here 
is the substantial suppression of the eigenvalue from "$(\pi,0)$" by 
including the intrapocket scattering processes; 
see the line of "intra + ($\pi$,0)" in \fig{lambda-T}~(b).  
Intrapocket scattering processes are characterized by small momentum transfers  
and correspond to a tail of spin fluctuations with a peak around $(\pi,0)$ and $(0,\pi)$. 
In fact, "intra" scattering processes alone yield  the eigenvalue less than 0.1. 
Therefore the contribution from "intra" scattering processes seems irrelevant to SC, 
but \fig{lambda-T}~(b) reveals that it   plays a vital role to suppress the SC tendency, 
which is the major finding of this work. 

This self-restraint effect can be understood in terms of phase frustration of pairing gap. 
As is well known \cite{scalapino12}, spin fluctuations give rise to a repulsive pairing interaction and in fact 
$\Gamma_{\vk_{F} \vk_{F}'} (i k_n, i k_n')$ in \eq{averagedGamma} is positive. 
In this case,  pairing gap tends to have the opposite sign between the hole and electron pockets connected 
by "$(\pi,0)$" scattering processes. The resulting gap 
has the same sign inside each pocket. On the other hand, spin fluctuations 
necessarily contain  "intra" scattering processes as a tail of the major antiferromagnetic fluctuations.  
These processes also yield a repulsive pairing interaction  and thus 
tend to drive the sign change of pairing gap {\it inside} each pocket. Therefore there occurs 
frustration of the phase of pairing gap from "$(\pi,0)$" and "intra" scattering processes. 
Figure~\ref{lambda-T}~(b) implies that this phase frustration effect 
is crucially important to the suppression of the eigenvalue of the Eliashberg equations 
even though the "intra" scattering processes alone are not effective to the SC instability itself. 
This self-restraint effect can be a general feature because 
the phase frustration is necessarily involved in the spin-fluctuation mechanism 
as long as it yields a repulsive pairing interaction. 

While "intra" scattering processes are the major source 
of the self-restraint effect, $"(\pi,\pi)"$ scattering processes also lead to the phase frustration 
of the SC gap. This is because they wish to have the {\it opposite} sign between the hole (electron) 
pockets whereas the major $"(\pi,0)"$ scattering processes eventually lead to the {\it same }sign 
between the hole (electron) pockets. 
Quantitatively, however, such a phase frustration effect is not effective 
compared to the "intra" processes as shown in \fig{lambda-T}~(b). 
In fact, the eigenvalue of the Eliashberg equations is almost reproduced by 
considering only "intra" and "$(\pi,0)$" scattering processes. 
That is, "intra" processes are much more destructive to the SC than "$(\pi, \pi)$" ones. 
For a different interaction $V(\vq)$, 
the contribution from $"(\pi,\pi)"$ scattering processes can suppress SC 
more than \fig{lambda-T}~(b), but still "intra" scattering processes play a major role 
of the self-restraint effect \cite{misc-supp-Lorentz}. 

The "intra" scattering processes should not be confused with ferromagnetic fluctuations. 
The self-restraint effect cannot be understood in terms of the competition of, 
for example, singlet and triplet pairings. 
In fact, the static magnetic susceptibility does not show any peak around $(0,0)$. 
Moreover, we checked that the eigenvector obtained from 
the "intra" pocket scattering processes alone is not triplet pairing. 

\begin{figure}[ht]
\centering
\includegraphics[width=12cm]{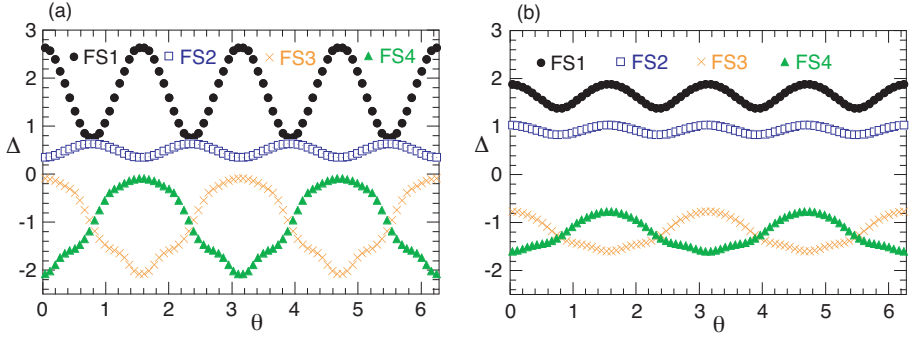}
\caption{(Color online)
Momentum dependence of the pairing gap $\Delta$ on each 
Fermi pocket at the lowest temperature $T=0.03$ from "all" scattering processes (a) 
and "$(\pi,0)$" scattering processes alone (b). The polar angle $\theta$ is 
measured from the horizontal axis on each pocket as shown in \fig{lambda-T}~(a). 
}
\label{gap-kf}
\end{figure}

To see how the self-restraint effect affects the momentum dependence 
 of the pairing gap, we plot $\vk_{F}$  dependence of the pairing gap in \fig{gap-kf} (Ref.~\onlinecite{misc-supp-Z}). 
 The pairing gap has the same sign in each pocket and the opposite sign between 
 the hole (FS1 and FS2) and 
 electron pockets (FS3 and FS4). The so-called $s_\pm$-wave symmetry is realized as expected \cite{mazin08,kuroki08}. 
 The pairing gap exhibits a large $\vk_{F}$ dependence on FS1, FS3, and FS4. 
 While the gap has a fourfold symmetry on FS1 and FS2, it has a two-fold symmetry on FS3 and FS3, 
 because the FS has a two-fold symmetry around $\vk=(\pi,0)$ and $(0,\pi)$, respectively. 
 All these features are consistent with the literature \cite{thomale11}.    
The point here is that those gaps suffer from the self-restraint effect. 
 The pairing gap without the self-restraint effect is 
 obtained by considering "$(\pi,0)$" scattering processes alone  
 and the obtained results are shown in \fig{gap-kf}~(b). 
 A comparison with \fig{gap-kf}~(a) demonstrates that 
 the self-restraint effect causes the large $\vk_{F}$ dependence 
 of the pairing gap on FS1, FS3, and FS4 to minimize the phase frustration effect 
 of the pairing gap although the $s_{\pm}$ symmetry does not change.

The self-restraint effect is different from the self-energy effect. 
We  compute the eigenvalue of the Eliashberg equations by 
neglecting the self-energy effect, namely by putting $Z=1$. 
The result is shown in \fig{lambda-T-Z1} in the same fashion as \fig{lambda-T}~(b) and 
essentially the same results are obtained except for the absolute value of $\lambda$. 
The $"(\pi,0)"$ scattering processes yield the SC instability 
at $T=0.042$, which is then reduced to $T=0.034$ by adding "intra" scattering processes; 
the resulting eigenvalue then reproduces the eigenvalue for "all" scattering processes. 
The self-restraint effect reduces $T_c$ by $(0.042-0.034)/0.042=19\%$. 
At $T=0.042$, we have obtained $\lambda=0.65$ in \fig{lambda-T} (b) for $"(\pi,0)"$ 
scattering processes. Hence the self-energy effect suppresses the SC tendency 
by $(1-0.65)/1=35\%$. That is, the suppression of the SC instability due to the self-restraint effect 
is comparable to that due to the self-energy effect.

Antiferromagnetic spin fluctuations are widely discussed as a possible 
high-$T_c$ mechanism. 
While there is no doubt that spin fluctuations can drive the SC, this mechanism needs to overcome 
the self-restraint effect to achieve high-$T_c$. In this sense, a favorable condition is 
required to realize high-$T_c$ from spin fluctuations. To reduce the self-restraint effect substantially, 
we would invoke an interaction term $V(\vq)$, whose magnitude becomes 
very small for a small momentum transfer 
so that the contribution from "intra" scattering processes is substantially weakened. 

On the other hand, orbital fluctuations with a large momentum 
transfer \cite{stanescu08,kontani11} and nematic fluctuations \cite{yamase13b,agatsuma16} 
are also proposed as a possible high-$T_c$ mechanism in FeSC. These fluctuations yield 
an attractive pairing interaction and thus tend to have the same sign of the paring gap on all FSs 
as far as we neglect the effect of spin fluctuations \cite{zhou11,yamada14}. 
Hence the self-restraint effect does not occur and all "intra", "$(\pi, 0)$", and  "$(\pi, \pi)$" 
scattering processes work positively for the SC instability. In this sense, it seems easier  
to achieve high-$T_c$ if those fluctuations are dominant. 
While the electron-phonon coupling is believed to be too small to explain $T_c$ of FeSC \cite{boeri08}, 
it is also free from the self-restraint effect 
as long as it yields an attractive pairing interaction.

In summary, it is tacitly assumed that antiferromagnetic spin fluctuations work positively 
for a SC instability. However, the present work finds that spin fluctuations have a 
contribution to {\it suppress} the SC tendency. 
This self-restraint effect comes from scattering processes inside the Fermi pockets 
with a small momentum transfer, which corresponds to a {\it tail} of the major antiferromagnetic spin fluctuations. 
We have shown that such a seemingly negligible contribution plays a remarkably important role 
to suppress the SC instability (Figs.~\ref{lambda-T} and \ref{lambda-T-Z1}). 
This effect is comparable to the suppression 
of SC by the self-energy effect. The self-restraint effect can be understood 
in terms of phase frustration of the paring gap caused by 
a repulsive pairing interaction inherent in antiferromagnetic spin fluctuations. 
To compromise with the frustration, the system tends to have a larger $\vk_{F}$ dependence 
of the pairing gap (\fig{gap-kf}).   
The self-restraint effect is general and thus expected also in other models of SC mediated by antiferromagnetic 
fluctuations \cite{scalapino12}. 

The authors thank W. Metzner for a critical reading of the manuscript and 
T. Hotta, A. Katanin, and T. Kita for fruitful comments. 
This work was supported by JSPS KAKENHI Grants No.~JP15K05189,  
JP18K18744. and JP20H01856.

\bibliography{main} 

\begin{thebibliography}{28}
\expandafter\ifx\csname natexlab\endcsname\relax\def\natexlab#1{#1}\fi
\expandafter\ifx\csname bibnamefont\endcsname\relax
  \def\bibnamefont#1{#1}\fi
\expandafter\ifx\csname bibfnamefont\endcsname\relax
  \def\bibfnamefont#1{#1}\fi
\expandafter\ifx\csname citenamefont\endcsname\relax
  \def\citenamefont#1{#1}\fi
\expandafter\ifx\csname url\endcsname\relax
  \def\url#1{\texttt{#1}}\fi
\expandafter\ifx\csname urlprefix\endcsname\relax\def\urlprefix{URL }\fi
\providecommand{\bibinfo}[2]{#2}
\providecommand{\eprint}[2][]{\url{#2}}

\bibitem[{\citenamefont{Stewart}(2011)}]{stewart11}
\bibinfo{author}{\bibfnamefont{G.~R.} \bibnamefont{Stewart}},
  \bibinfo{journal}{Rev. Mod. Phys.} \textbf{\bibinfo{volume}{83}},
  \bibinfo{pages}{1589} (\bibinfo{year}{2011}).

\bibitem[{\citenamefont{Mazin et~al.}(2008)\citenamefont{Mazin, Singh,
  Johannes, and Du}}]{mazin08}
\bibinfo{author}{\bibfnamefont{I.~I.} \bibnamefont{Mazin}},
  \bibinfo{author}{\bibfnamefont{D.~J.} \bibnamefont{Singh}},
  \bibinfo{author}{\bibfnamefont{M.~D.} \bibnamefont{Johannes}},
  \bibnamefont{and} \bibinfo{author}{\bibfnamefont{M.~H.} \bibnamefont{Du}},
  \bibinfo{journal}{Phys.\ Rev.\ Lett.} \textbf{\bibinfo{volume}{101}},
  \bibinfo{pages}{057003} (\bibinfo{year}{2008}).

\bibitem[{\citenamefont{Kuroki et~al.}(2008)\citenamefont{Kuroki, Onari, Arita,
  Usui, Tanaka, Kontani, and Aoki}}]{kuroki08}
\bibinfo{author}{\bibfnamefont{K.}~\bibnamefont{Kuroki}},
  \bibinfo{author}{\bibfnamefont{S.}~\bibnamefont{Onari}},
  \bibinfo{author}{\bibfnamefont{R.}~\bibnamefont{Arita}},
  \bibinfo{author}{\bibfnamefont{H.}~\bibnamefont{Usui}},
  \bibinfo{author}{\bibfnamefont{Y.}~\bibnamefont{Tanaka}},
  \bibinfo{author}{\bibfnamefont{H.}~\bibnamefont{Kontani}}, \bibnamefont{and}
  \bibinfo{author}{\bibfnamefont{H.}~\bibnamefont{Aoki}},
  \bibinfo{journal}{Phys.\ Rev.\ Lett.} \textbf{\bibinfo{volume}{101}},
  \bibinfo{pages}{087004} (\bibinfo{year}{2008}).

\bibitem[{\citenamefont{Chubukov et~al.}(2008)\citenamefont{Chubukov, Efremov,
  and Eremin}}]{chubukov08}
\bibinfo{author}{\bibfnamefont{A.~V.} \bibnamefont{Chubukov}},
  \bibinfo{author}{\bibfnamefont{D.~V.} \bibnamefont{Efremov}},
  \bibnamefont{and} \bibinfo{author}{\bibfnamefont{I.}~\bibnamefont{Eremin}},
  \bibinfo{journal}{Phys. Rev. B} \textbf{\bibinfo{volume}{78}},
  \bibinfo{pages}{134512} (\bibinfo{year}{2008}).

\bibitem[{\citenamefont{Fernandes et~al.}(2014)\citenamefont{Fernandes,
  Chubukov, and Schmalian}}]{fernandes14}
\bibinfo{author}{\bibfnamefont{R.~M.} \bibnamefont{Fernandes}},
  \bibinfo{author}{\bibfnamefont{A.~V.} \bibnamefont{Chubukov}},
  \bibnamefont{and}
  \bibinfo{author}{\bibfnamefont{J.}~\bibnamefont{Schmalian}},
  \bibinfo{journal}{Nat. Phys.} \textbf{\bibinfo{volume}{10}},
  \bibinfo{pages}{97} (\bibinfo{year}{2014}).

\bibitem[{\citenamefont{Yamase and Zeyher}(2013)}]{yamase13b}
\bibinfo{author}{\bibfnamefont{H.}~\bibnamefont{Yamase}} \bibnamefont{and}
  \bibinfo{author}{\bibfnamefont{R.}~\bibnamefont{Zeyher}},
  \bibinfo{journal}{Phys. Rev. B} \textbf{\bibinfo{volume}{88}},
  \bibinfo{pages}{180502(R)} (\bibinfo{year}{2013}).

\bibitem[{\citenamefont{Agatsuma and Yamase}(2016)}]{agatsuma16}
\bibinfo{author}{\bibfnamefont{T.}~\bibnamefont{Agatsuma}} \bibnamefont{and}
  \bibinfo{author}{\bibfnamefont{H.}~\bibnamefont{Yamase}},
  \bibinfo{journal}{Phys. Rev. B} \textbf{\bibinfo{volume}{94}},
  \bibinfo{pages}{214505} (\bibinfo{year}{2016}).

\bibitem[{\citenamefont{Stanescu et~al.}(2008)\citenamefont{Stanescu, Galitski,
  and Sarma}}]{stanescu08}
\bibinfo{author}{\bibfnamefont{T.~D.} \bibnamefont{Stanescu}},
  \bibinfo{author}{\bibfnamefont{V.}~\bibnamefont{Galitski}}, \bibnamefont{and}
  \bibinfo{author}{\bibfnamefont{S.~D.} \bibnamefont{Sarma}},
  \bibinfo{journal}{Phys.\ Rev.\ B} \textbf{\bibinfo{volume}{78}},
  \bibinfo{pages}{195114} (\bibinfo{year}{2008}).

\bibitem[{\citenamefont{Kontani et~al.}(2011)\citenamefont{Kontani, Saito, and
  Onari}}]{kontani11}
\bibinfo{author}{\bibfnamefont{H.}~\bibnamefont{Kontani}},
  \bibinfo{author}{\bibfnamefont{T.}~\bibnamefont{Saito}}, \bibnamefont{and}
  \bibinfo{author}{\bibfnamefont{S.}~\bibnamefont{Onari}},
  \bibinfo{journal}{Phys. Rev. B} \textbf{\bibinfo{volume}{84}},
  \bibinfo{pages}{024528} (\bibinfo{year}{2011}).

\bibitem[{mis({\natexlab{a}})}]{misc-orbital}
\bibinfo{note}{We distinguish between orbital nematic fluctuations and orbital
  fluctuations because the underlaying spectra are different: the dominant
  fluctuations occur around $\vq=(0,0)$ in the former whereas the latter
  accompanies a large momentum transfer.}

\bibitem[{\citenamefont{Boeri et~al.}(2008)\citenamefont{Boeri, Dolgov, and
  Golubov}}]{boeri08}
\bibinfo{author}{\bibfnamefont{L.}~\bibnamefont{Boeri}},
  \bibinfo{author}{\bibfnamefont{O.~V.} \bibnamefont{Dolgov}},
  \bibnamefont{and} \bibinfo{author}{\bibfnamefont{A.~A.}
  \bibnamefont{Golubov}}, \bibinfo{journal}{Phys. Rev. Lett.}
  \textbf{\bibinfo{volume}{101}}, \bibinfo{pages}{026403}
  (\bibinfo{year}{2008}).

\bibitem[{\citenamefont{Yanagi et~al.}(2010)\citenamefont{Yanagi, Yamakawa,
  Adachi, and \={O}no}}]{yanagi10}
\bibinfo{author}{\bibfnamefont{Y.}~\bibnamefont{Yanagi}},
  \bibinfo{author}{\bibfnamefont{Y.}~\bibnamefont{Yamakawa}},
  \bibinfo{author}{\bibfnamefont{N.}~\bibnamefont{Adachi}}, \bibnamefont{and}
  \bibinfo{author}{\bibfnamefont{Y.}~\bibnamefont{\={O}no}},
  \bibinfo{journal}{Phys. Rev. B} \textbf{\bibinfo{volume}{82}},
  \bibinfo{pages}{064518} (\bibinfo{year}{2010}).

\bibitem[{\citenamefont{Yamada et~al.}(2014)\citenamefont{Yamada, Ishizuka, and
  \={O}no}}]{yamada14}
\bibinfo{author}{\bibfnamefont{T.}~\bibnamefont{Yamada}},
  \bibinfo{author}{\bibfnamefont{J.}~\bibnamefont{Ishizuka}}, \bibnamefont{and}
  \bibinfo{author}{\bibfnamefont{Y.}~\bibnamefont{\={O}no}},
  \bibinfo{journal}{J. Phys. Soc. Jpn.} \textbf{\bibinfo{volume}{83}},
  \bibinfo{pages}{043704} (\bibinfo{year}{2014}).

\bibitem[{\citenamefont{{S. Raghu, X.-L. Qi, C.-X. Liu, D. J. Scalapino, and
  S.-C. Zhang}}(2008)}]{raghu08}
\bibinfo{author}{\bibnamefont{{S. Raghu, X.-L. Qi, C.-X. Liu, D. J. Scalapino,
  and S.-C. Zhang}}}, \bibinfo{journal}{Phys.\ Rev.\ B}
  \textbf{\bibinfo{volume}{77}}, \bibinfo{pages}{220503}
  (\bibinfo{year}{2008}).

\bibitem[{\citenamefont{{Z.-J. Yao, J.-X. Li, and Z. D. Wang}}(2009)}]{yao09}
\bibinfo{author}{\bibnamefont{{Z.-J. Yao, J.-X. Li, and Z. D. Wang}}},
  \bibinfo{journal}{New J. Phys.} \textbf{\bibinfo{volume}{11}},
  \bibinfo{pages}{025009} (\bibinfo{year}{2009}).

\bibitem[{\citenamefont{Graser et~al.}(2009)\citenamefont{Graser, Maier,
  Hirschfeld, and Scalapino}}]{graser09}
\bibinfo{author}{\bibfnamefont{S.}~\bibnamefont{Graser}},
  \bibinfo{author}{\bibfnamefont{T.~M.} \bibnamefont{Maier}},
  \bibinfo{author}{\bibfnamefont{P.~J.} \bibnamefont{Hirschfeld}},
  \bibnamefont{and} \bibinfo{author}{\bibfnamefont{D.~J.}
  \bibnamefont{Scalapino}}, \bibinfo{journal}{New J. Phys.}
  \textbf{\bibinfo{volume}{11}}, \bibinfo{pages}{025016}
  (\bibinfo{year}{2009}).

\bibitem[{\citenamefont{Husemann and Salmhofer}(2009)}]{husemann09}
\bibinfo{author}{\bibfnamefont{C.}~\bibnamefont{Husemann}} \bibnamefont{and}
  \bibinfo{author}{\bibfnamefont{M.}~\bibnamefont{Salmhofer}},
  \bibinfo{journal}{Phys. Rev. B} \textbf{\bibinfo{volume}{79}},
  \bibinfo{pages}{195125} (\bibinfo{year}{2009}).

\bibitem[{\citenamefont{Eberlein and Metzner}(2014)}]{eberlein14}
\bibinfo{author}{\bibfnamefont{A.}~\bibnamefont{Eberlein}} \bibnamefont{and}
  \bibinfo{author}{\bibfnamefont{W.}~\bibnamefont{Metzner}},
  \bibinfo{journal}{Phys. Rev. B} \textbf{\bibinfo{volume}{89}},
  \bibinfo{pages}{035126} (\bibinfo{year}{2014}).

\bibitem[{\citenamefont{Dai et~al.}(2012)\citenamefont{Dai, Hu, and
  Dagotto}}]{dai12}
\bibinfo{author}{\bibfnamefont{P.}~\bibnamefont{Dai}},
  \bibinfo{author}{\bibfnamefont{J.}~\bibnamefont{Hu}}, \bibnamefont{and}
  \bibinfo{author}{\bibfnamefont{E.}~\bibnamefont{Dagotto}},
  \bibinfo{journal}{Nat. Phys.} \textbf{\bibinfo{volume}{8}},
  \bibinfo{pages}{709} (\bibinfo{year}{2012}).

\bibitem[{mis({\natexlab{b}})}]{misc-supp-Lorentz}
\bibinfo{note}{See Supplemental Material at [URL] for results of the
  Lorentz-type interaction $V(\bf{q})$.}

\bibitem[{\citenamefont{Schrieffer}(1999)}]{schrieffer}
\bibinfo{author}{\bibfnamefont{J.~R.} \bibnamefont{Schrieffer}},
  \emph{\bibinfo{title}{Theory of Superconductivity}}
  (\bibinfo{publisher}{Perseus Books}, \bibinfo{year}{1999}),
  \bibinfo{edition}{revised} ed.

\bibitem[{\citenamefont{Arita and Ikeda}(2009)}]{arita09}
\bibinfo{author}{\bibfnamefont{R.}~\bibnamefont{Arita}} \bibnamefont{and}
  \bibinfo{author}{\bibfnamefont{H.}~\bibnamefont{Ikeda}}, \bibinfo{journal}{J.
  Phys. Soc. Jpn.} \textbf{\bibinfo{volume}{78}}, \bibinfo{pages}{113707}
  (\bibinfo{year}{2009}).

\bibitem[{\citenamefont{Suzuki et~al.}(2014)\citenamefont{Suzuki, Usui, Iimura,
  Sato, Matsuishi, Hosono, and Kuroki}}]{suzuki14}
\bibinfo{author}{\bibfnamefont{K.}~\bibnamefont{Suzuki}},
  \bibinfo{author}{\bibfnamefont{H.}~\bibnamefont{Usui}},
  \bibinfo{author}{\bibfnamefont{S.}~\bibnamefont{Iimura}},
  \bibinfo{author}{\bibfnamefont{Y.}~\bibnamefont{Sato}},
  \bibinfo{author}{\bibfnamefont{S.}~\bibnamefont{Matsuishi}},
  \bibinfo{author}{\bibfnamefont{H.}~\bibnamefont{Hosono}}, \bibnamefont{and}
  \bibinfo{author}{\bibfnamefont{K.}~\bibnamefont{Kuroki}},
  \bibinfo{journal}{Phys. Rev. Lett.} \textbf{\bibinfo{volume}{113}},
  \bibinfo{pages}{027002} (\bibinfo{year}{2014}).

\bibitem[{\citenamefont{Usui et~al.}(2015)\citenamefont{Usui, Suzuki, and
  Kuroki}}]{usui15}
\bibinfo{author}{\bibfnamefont{H.}~\bibnamefont{Usui}},
  \bibinfo{author}{\bibfnamefont{K.}~\bibnamefont{Suzuki}}, \bibnamefont{and}
  \bibinfo{author}{\bibfnamefont{K.}~\bibnamefont{Kuroki}},
  \bibinfo{journal}{Scientific Reports} \textbf{\bibinfo{volume}{5}},
  \bibinfo{pages}{11399} (\bibinfo{year}{2015}).

\bibitem[{\citenamefont{Scalapino}(2012)}]{scalapino12}
\bibinfo{author}{\bibfnamefont{D.~J.} \bibnamefont{Scalapino}},
  \bibinfo{journal}{Rev. Mod. Phys.} \textbf{\bibinfo{volume}{84}},
  \bibinfo{pages}{1383} (\bibinfo{year}{2012}).

\bibitem[{mis({\natexlab{c}})}]{misc-supp-Z}
\bibinfo{note}{See Supplemental Material at [URL] for the momentum dependence
  of the renormalization function.}

\bibitem[{\citenamefont{Thomale et~al.}(2011)\citenamefont{Thomale, Platt,
  Hanke, and Bernevig}}]{thomale11}
\bibinfo{author}{\bibfnamefont{R.}~\bibnamefont{Thomale}},
  \bibinfo{author}{\bibfnamefont{C.}~\bibnamefont{Platt}},
  \bibinfo{author}{\bibfnamefont{W.}~\bibnamefont{Hanke}}, \bibnamefont{and}
  \bibinfo{author}{\bibfnamefont{B.~A.} \bibnamefont{Bernevig}},
  \bibinfo{journal}{Phys. Rev. Lett.} \textbf{\bibinfo{volume}{106}},
  \bibinfo{pages}{187003} (\bibinfo{year}{2011}).

\bibitem[{\citenamefont{Zhou et~al.}(2011)\citenamefont{Zhou, Kotliar, and
  Wang}}]{zhou11}
\bibinfo{author}{\bibfnamefont{S.}~\bibnamefont{Zhou}},
  \bibinfo{author}{\bibfnamefont{G.}~\bibnamefont{Kotliar}}, \bibnamefont{and}
  \bibinfo{author}{\bibfnamefont{Z.}~\bibnamefont{Wang}},
  \bibinfo{journal}{Phys. Rev. B} \textbf{\bibinfo{volume}{84}},
  \bibinfo{pages}{140505(R)} (\bibinfo{year}{2011}).

\end{thebibliography}


\newpage
\begin{center}
{\bf SUPPLEMENTAL MATERIAL}
\end{center}

\section{Momentum dependence of renormalization function} 
\begin{figure}[th]
\centering
\includegraphics[width=6cm]{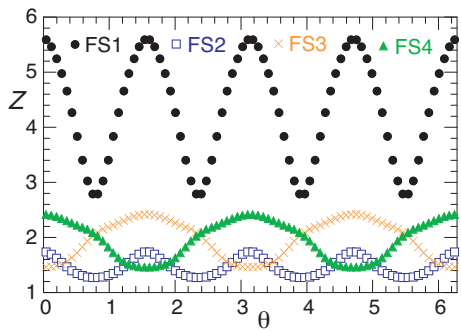}
\caption{(Color online)
Momentum dependence of the renormalization function $Z$ on each 
Fermi pockets at the lowest temperature $T=0.03$. The polar angle $\theta$ is 
measured from the horizontal axis on each pocket as shown in \fig{lambda-T}~(a). 
}
\label{Z-kf}
\end{figure}

We show in Fig.~\ref{Z-kf} the momentum dependence of the renormalization function $Z$ 
along the FSs at the lowest temperature; 
the corresponding results of $\lambda$ and $\Delta$ are shown in  Figs.~\ref{lambda-T}~(b) 
and \ref{gap-kf}~(a), respectively. 
While a value of $Z$ on FS2 stays in 1.2 - 1.8 and thus weak-coupling theory 
would work there, the value of  $Z$ amounts to 
5.6 on FS1 and 2.4 on FS3 and FS4, 
indicating the importance of the self-energy effect beyond the weak coupling theory. 

\section{Lorentz-type interaction} 
In the main text we have considered the magnetic interaction, which extends up to 
the second nearest-neighbor sites in real space. 
As the {\it opposite} limit, one may consider the Lorentz-type interaction 
describing an exponential-like decay in real space, 
\be
V_{L} (\vq) = -2 V \sum_{l=1}^{2} 
\sum_{n, m} \frac{\Gamma}{(\vq - \vQ^{n m}_{l})^{2} + \Gamma^2} \,,
\label{Lorentz} 
\ee
where $\vQ_{1}^{n m}=(\pi+ 2n\pi, 2m\pi)$, $\vQ_{2}^{n m}=(2 m \pi, \pi+ 2n\pi)$,  
and $n$ and $m$ are integers. 
$\Gamma$ determines the peak width and $V$ the magnitude. 

\begin{figure}[ht]
\centering
\includegraphics[width=7cm]{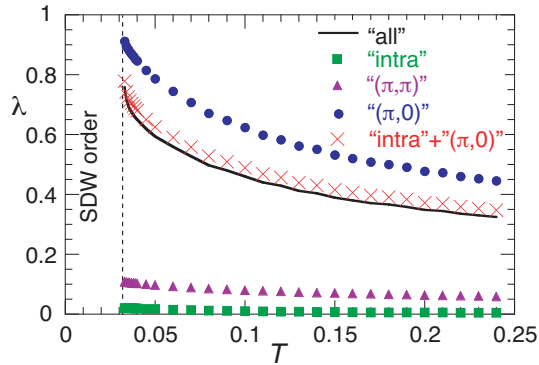}
\caption{(Color online)
Temperature dependence of the eigenvalues $\lambda$ 
(solid line denoted by "all") for the Lorentz-type magnetic 
interaction. The eigenvalues are also computed by focusing on 
particular scattering processes as denoted by 
"intra", "$(\pi,\pi)$", and "$(\pi,0)$". 
Below $T=0.032$, SDW order occurs before SC instability. 
}
\label{lambda-T-L}
\end{figure}

We perform the same calculations as \fig{lambda-T}~(b), 
but employing the interaction term \eq{Lorentz}. We take the parameters as 
$n=0, \pm 1, \pm 2, \pm 3, -4$, $m=0, \pm 1, \pm 2, \pm 3$, $\Gamma=1$, 
and $V=2.1$ for which the system has a SDW long range order below $T=0.032$. 
Obtained results are shown in \fig{lambda-T-L}, which is essentially the same as \fig{lambda-T}~(b).  
First, the eigenvalue of the Eliashberg equations is determined practically by 
the scattering processes "intra+$(\pi,0)$". Second, the SC tendency 
from "$(\pi,0)$" scattering processes is substantially suppressed by "intra" scattering processes. 
This self-restraint effect is weaker than \fig{lambda-T}~(b). 
This can be easily understood by observing in \fig{lambda-T-L} that "intra" scattering processes alone 
yield eigenvalues smaller than those in \fig{lambda-T}~(b) and thus the effect of "intra" scattering 
processes should become weaker. 
On the other hand, 
the $"(\pi,\pi)"$ scattering processes alone give eigenvalues larger than those in \fig{lambda-T}~(b). 
However, the eigenvalue of 
the Eliashberg equations is almost reproduced by considering the "intra+$(\pi,0)$" 
scattering processes only. In this sense, "intra" scattering processes are much more 
destructive to the SC than "$(\pi,\pi)$" scattering processes.  

\end{document}